\documentclass[prd,superscriptaddress,twocolumn,showpacs,preprintnumbers,amsmath,amssymb,floatfix]{revtex4}
\usepackage{graphicx}
\usepackage{dcolumn}
\usepackage{latexsym}
\usepackage{psfig}
\newcommand{\beq}{\begin{equation}}
\newcommand{\eeq}{\end{equation}}
\newcommand{\beqn}{\begin{eqnarray}}
\newcommand{\eeqn}{\end{eqnarray}}
\newcommand{\bea}[1]{\beq\begin{array}{#1}}
\newcommand{\eea}{\end{array}\eeq}
\newcommand{\eq}[1]{(\ref{#1})}

\newcommand{\tr}{\mathop{\rm Tr}}

\newcommand{\Pexp}{\mbox{P}\!\exp}

\newcommand{\ket}[1]{|\,#1\,\rangle}
\newcommand{\bra}[1]{\langle\,#1\,|}
\newcommand{\braket}[2]{\langle\,#1\,|\,#2\,\rangle}

\newcommand{\dd}{{\mathrm d}}
\newcommand{\Z}{Z\!\!\! Z}

\newcommand{\LQ}{\Lambda_{QCD}}
\newcommand{\diff}{\partial}

\newcommand{\cC}{{\cal C}}

\newcommand{\cD}{{\cal D}}

\newcommand{\HP}[1]{{\mathrm{HP}^#1}}
\newcommand{\hp}{\mathrm{HP}}

\newcommand{\mean}[1]{{\langle #1 \rangle}}

\newcommand{\Hiroshima}{\affiliation{RIISE, Hiroshima University, Higashi-Hiroshima, 739-8527,
Japan}}
\newcommand{\ITEP}{\affiliation{ITEP, B.Cheremushkinskaya 25, Moscow, 117218, Russia}}

\begin{document}
\preprint{ITEP-LAT/2007-03}

\title{Confining String and Its Widening in $\HP{1}$ Embedding Approach}

\author{M.~N.~Chernodub}\ITEP\Hiroshima
\author{F.~V.~Gubarev}\ITEP

\begin{abstract}
Structure of confining string in terms of the topological charge density and the action density
is studied in $SU(2)$  Yang-Mills theory on the lattice using $\HP{1}$ $\sigma$-model embedding
approach. We find that the confining flux tube noticeably suppresses both
the topological charge and the action densities. Beyond the string formation length
the string cross section in terms of these quantities is well described by
a Gaussian profile.
In both cases the squared string width is found to be a logarithmic function of the string
length confirming the L\"uscher widening of the chromoelectric string.
Characteristic string scales in terms of the topological and action densities are estimated as well.
\end{abstract}


\maketitle

\section{Introduction}

It is commonly believed that the phenomenon of the quark confinement in QCD happens due to formation
of the chromoelectric string spanned between quark and anti-quark. In fact, various numerical
simulations in QCD both in quenched and in unquenched limits indicate that in the QCD vacuum the
fundamental color charges (quarks) are connected by a string-like structure~\cite{Bali:1994de} (see,
e.g., Ref~\cite{Bali:review} and references therein), which can be seen directly in gluonic energy
and action density~\cite{ref:string:energy-action,Bissey} as well as in other related
quantities~\cite{ref:string:other}. Physically, the string is formed by squeezed fluxes of
chromoelectric fields coming out of quarks. The chromoelectric string possesses a non-zero tension
providing steady attracting force which makes the quarks confined into colorless hadrons.

There are various effects caused by the strong chromoelectric fields of the QCD string.
Besides the enhancement of the gluonic energy density, the string is characterized by the
suppression of the gluonic action density~\cite{ref:sum-rules},
as well as by the suppression~\cite{Chernodub:2005gz} of the $\mean{A^2}$-condensate~\cite{ref:A2}.
Some features of the QCD string are consistent with the dual superconductor
model of the color confinement~\cite{ref:dual:superconductor} which attributes
the color flux squeezing to a (dual) Meissner
effect originating from condensation of particular (``monopole-like'') gluon field
configurations~\cite{ref:reviews}.
Naturally, chromoelectric string formation has a great impact on both gluons and quarks,
however, in latter case  the confinement of color manifests itself in yet another way.
There are numerical indications that the chiral symmetry, originally broken
by non-perturbative effects in the low temperature QCD, is (partially) restored inside the
string~\cite{ref:Laermann}. The effect is revealed by local suppression of the
chiral condensate $\langle \bar \psi \psi\rangle$ in the vicinity of the string.
The same effect is also observed numerically in the neighborhood of individual heavy
quarks~\cite{ref:chiral:quark}, indicating that the strong chromoelectric fields
tend to restore the chiral symmetry.
Since  the chiral properties of the vacuum are tightly related to its topology,
one expects generically that  strong chromoelectric field of the string
severely affects the topological charge density.
Indeed, the anti-correlation between the QCD string and the topological charge density was
discovered in numerical simulation of Ref.~\cite{string:topcharge:first} in the so
called ``cooled'' $SU(3)$ vacuum. The investigation was developed further in Ref.~\cite{string:topcharge}
with the help of the same cooling method which is a gradual transformation of
the quantum (``hot'') fields towards the classical (``cold'') vacuum. The cooling is aimed
to reduce the topologically trivial effects of divergent zero-point fluctuations.
This important result -- obtained with the simplest lattice definitions of the
topological charge -- is a first numerical indication that the QCD string
suppresses topological fluctuations.
In this paper we investigate, in particular, the structure of confining string in terms of
the topological charge density.  The problem is conceptually and technically challenging
and  therefore it calls for a few physically motivated simplifications. First,
we refrain from using the cooling approach because it certainly wipes out local topology
of the real (uncooled) vacuum. Indeed,  the fluctuations of the topological charge
inside the chromoelectric string were found to depend crucially upon the cooling
parameters~\cite{string:topcharge:first,string:topcharge} and therefore are affected by uncontrollable systematic errors.
Other essential simplifications used in this paper are as follows:

{\it Quenched vacuum.}
For our purposes it is natural  to
consider
quenched approximation because dynamical quarks screen
the test color charges and break the string. As a result the correlation
between the string and any locally defined quantity disappears at sufficiently
large separations between the test charges. Thus the choice of the quenched vacuum
is crucial,  as  it allows to discriminate between the string breaking and
the actual correlation effects.

{\it Reduced number of colors.}
It is advantageous to utilize the known similarity between the $SU(2)$ and $SU(3)$ gauge models.
Since nonperturbative aspects are alike in these theories, we naturally expect that the
behavior of the topological charge density around the string in the realistic $SU(3)$ case is
qualitatively the same as for the $SU(2)$ gauge group studied in our paper.
Apart from the obvious computational simplifications, restriction to $SU(2)$ gauge theory
allows us to address unambiguously the next urgent problem of

{\it Ultraviolet (UV) fluctuations.}
Indeed, simply on dimensional grounds one expects that the vacuum topological density
is power-like divergent $\mean{q^2} \sim 1/a^8$, $a$ being the lattice spacing (the inverse UV cutoff).
However, the leading UV divergence should not be sensitive to the quark-antiquark separation.
Note that this is not so in the case of subleading power corrections, which
depend also upon the infrared (IR) scale (denoted $\LQ$ generically).
Therefore  in order to study the correlation of the string and the topological density it would be
advantageous to remove the leading (and only leading) power divergence in
$\mean{q^2}$ since it gives identically vanishing but otherwise numerically
destructive  contribution to the correlation function.

In view of the above points  the choice of the topological density operator,
which is free from additional parameters, becomes, in fact, almost unique.
Indeed, the cooling approach was rejected for reasons explained above.
As far as the fermionic definition~\cite{overlap}
is concerned, in the numerical implementations it is either not parameter free or is not capable
to filter out exclusively the leading power divergence.
Indeed, the essence of the construction is to sum up the individual contributions
of eigenmodes of overlap Dirac operator, ranked with corresponding eigenvalues.
It is known~\cite{Horvath:2005cv}
that the full tower of eigenmodes indeed reproduces $\sim 1/a^8$ behavior, while
the restriction to lowest eigenvalues~\cite{Horvath:2003yj} introduces explicit
and {\it a priori} ambiguous cut on the leading singularity.
Therefore the essentially unique alternative construction which fits the above requirements
is $\HP{1}$ $\sigma$-model embedding approach~\cite{Gubarev:2005rs,Boyko:2006jt}
which is to be reviewed in Section~~\ref{sec:HP1}.
Here we only mention that this method indeed provides parameter free definition of the
topological density in which the leading power divergence is removed.
Note that the removal of leading perturbative contributions from various observables
in $\HP{1}$ embedding approach had not been proved rigorously, however, in all examples
considered so far it turns out to be the case. Moreover, the subleading power corrections
seem to be left intact by the corresponding gauge covariant $\HP{1}$-projection procedure.
What is even more important is that near the continuum limit the $\HP{1}$-projected fields
capture the majority of non-perturbative aspects of $SU(2)$ Yang-Mills theory
and reproduce, for instance, the full tension of the confining string and
(quenched) quark condensate (see Refs.~\cite{Gubarev:2005rs,Boyko:2006jt} for further details).
Being numerically superior, the $\HP{1}$ embedding approach is naturally
our method of choice to investigate the topological aspects of the confining string
and to study its geometrical properties. For discussion of other methods which allow
to ``filter out'' topological properties of the vacuum out of the ultraviolet noise see
Ref.~\cite{ref:other:methods}.

However, the advantages of our method are not coming for free, the price
is the intrinsic non-locality: the $\HP{1}$ $\sigma$-model fields
constructed from given $SU(2)$ gauge background  are not local in terms
of the original gauge potentials. However the characteristic range of non-locality is
certainly less than $\lesssim 0.3 ~\mathrm{fm}$
and is comparable with the expected~\cite{Bali:1994de} half-width of the flux tube
as it is determined by the string action density profile.
We conclude therefore that the non-locality of our method should not be essential for
large enough quark-antiquark separations.

The knowledge of the string profiles immediately reveals an
important feature of the chromoelectric string known as the string
widening. This rather general phenomenon occurs in various gauge and spin systems
possessing the string-like excitations. As it was predicted
by L\"uscher~\cite{Luscher:1980iy}, quantum fluctuations force the string to fluctuate
transversely in such a way that its geometrical center follows the Gaussian distribution.
The squared width of the Gaussian distribution (string width)
is expected to be the logarithmic function of the string length, $\delta^2(R) \propto \ln R$,
which in quenched QCD is the distance between heavy quark-antiquark pair.

The logarithmic widening of the string was accurately seen~\cite{Caselle:1995fh}
in the $\Z_2$ gauge model in $(2+1)$ dimensions, and the signatures of the widening are reflected
in the form of the potential between test sources connected by the string~\cite{Panero:String:Effects}.
For other gauge theories the results
are still inconclusive. For instance, the data for the string in gluodynamics width obtained via
the energy density profiles~\cite{Bali:1994de} is consistent
both with the constant behavior, $\delta(R) \approx {\mathrm{const}}$, and with
the logarithmic scaling for $R \gtrsim 1\,\mbox{fm}$.

In the numerical simulations the string profile is usually overshadowed
by the ultraviolet noise, which may be reduced within the certain models
of the vacuum structure. For instance, the dual superconductor scenario of quark
confinement~\cite{ref:dual:superconductor} suggests that the confining string formation
is governed by the dynamics of Abelian fields calculated in a certain Abelian gauge.
While this approach consistently describes particular features of the chromoelectric
string, the string width is found to be length-independent both in quenched~\cite{Bali:review}
and full~\cite{Bornyakov:2003vx} QCD, thus contradicting the logarithmic scaling law.
Another popular approach~\cite{center:confinement} assumes that basic non-perturbative
properties of gluons are encoded in the center degrees of freedom,
the corresponding string profile was calculated in Ref.~\cite{Bornyakov:2003gn}.
It was found that the string width itself (not its square) is a logarithmic function
of the quark-antiquark distance, again contradicting the L\"uscher law.
Therefore the results available  in the literature indicate that neither Abelian nor
center degrees of freedom preserve the transverse non-perturbative fluctuations
of the confining string, while the simulations within the full theory give
inconclusive results.
In this paper we investigate the string broadening using $\HP{1}$-projected fields,
which allow us to obtain a clean picture of the confining string
both in terms of the topological density and in terms of the gluonic action distribution.

The structure of the paper is as follows. The essentials of $\HP{1}$ $\sigma$-model embedding
approach are reviewed in Sections~\ref{sec:HP1}, while in Section~\ref{sec:observables}
we discuss the relevant observables and theoretical expectations.
In Section~\ref{sec:numerics} we describe the technical details of our measurements.
Then the profiles of the QCD string in terms of the topological and $\HP{1}$-projected action
densities are presented in Sections~\ref{sec:topology},\ref{sec:action}, respectively, where
we also discuss the degree of  the string broadening at various quark-antiquark separations.
Section \ref{sec:discussion} is devoted to the discussion of our results.

\section{Review of $\HP{1}$ embedding}
\label{sec:HP1}

In this Section we briefly review the $\HP{N}$ $\sigma$-model embedding approach
using continuum notation, for precise lattice definitions and further
details see Refs.~\cite{Gubarev:2005rs,Boyko:2006jt}.
The most straightforward way to introduce the idea of the method is to consider first
Yang-Mills theory with gauge group $SU(2)$ (group of unit quaternions)
in the limit, in which the gauge potentials $A_\mu$ are smooth functions of space-time coordinates.
Then it is known~\cite{theorem}
(see, e.g., Ref.~\cite{theorem:review} for extended discussions) that $A_\mu$
can  be represented as
\beq
\label{HP1:theorem}
A_\mu = \bra{q} \diff_\mu \ket{q} = \sum_{i=0}^N \bar{q}_i \diff_\mu q_i\,,
\eeq
where $\ket{q}$ is $(N+1)\times 1$ normalized $\braket{q}{q} = 1$ column vector with quaternion valued
entries $q_i$, bar denotes quaternionic conjugation and $N$ is finite integer number, the upper bound on which
is unimportant for the present discussion.
Gauge covariance of \eq{HP1:theorem} implies that $\ket{q} = [q_0,...,q_N]^T$
and $\ket{q} v = [q_0 v,...,q_N v]^T$,
where $v$ is unit quaternion, $\bar{v} v = 1$, are physically equivalent and therefore the set of
gauge inequivalent fields $\ket{q}$ describe the quaternionic projective space $\HP{N}$, convenient
parametrization of which is provided by gauge invariant $(N+1)\times (N+1)$ projection matrices
$P = \ket{q}\bra{q}$.  Therefore in the limit of smooth fields the $SU(2)$ Yang-Mills theory could
be reformulated in terms of the non-linear $\sigma$-model with target space $\HP{N}$.
Note that in the particular case $N=1$ target space geometry simplifies greatly, $\HP{1} = S^4$, and one has
\beq
\label{HP1:projector}
P =  \ket{q}\bra{q} = \frac{1}{2} (1 + \gamma^A n^A)\,, \quad A = 1,...,5\,,
\eeq
where $n^A$ is unit five-dimensional vector and $\gamma^A$ are the conventional Dirac matrices.
We use the same symbol $q$ to denote both the quaternions and the topological
charge density.
We do not expect a confusion with notation since the quaternions are used only in this Section and they are
always associated with the bra-ket symbols.

The essence of $\HP{N}$ $\sigma$-model embedding approach is to consider Eq.~\eq{HP1:theorem} within
the Euclidean formulation of quantum $SU(2)$ Yang-Mills theory and to quantify its approximate
validity for given gauge background. Namely, for any fixed gauge configuration $A_\mu$ one  finds the ``nearest''
$\HP{N}$ $\sigma$-model fields $\{\ket{q_x}\}$, which minimize the functional norm
\beqn
\label{HP1:norm}
& || A - \bra{q} \diff \ket{q}|| \equiv F(A, \ket{q}) < F(A, \ket{q'})\,, & \\
& \forall \,\, \{\ket{q_x'}\} \ne \{\ket{q_x}\}\,. & \nonumber
\eeqn
In fact, the problem \eq{HP1:norm} is well suited for the lattice studies and it turns out that
the minimizing configuration $\{\ket{q_x}\}$ is essentially unique, at least for the gauge potentials numerically
generated on the lattice near the continuum limit.
Note that the only free parameter at this stage is the rank $N$ of the $\sigma$-model considered.
Naively, one could expect that the quality of approximation $A \approx \bra{q} \diff \ket{q}$
quantified by $F(A, \ket{q})$ varies strongly with $N$, however, this turns out not to be the case.
To the contrary, the functional distance $F(A, \ket{q})$ appears to be independent
on the $\sigma$-model rank and hence there are no practical reasons to go beyond the simplest $N=1$ case.
In turn the restriction to $N=1$ allows to give natural and computationally superior
definition of the topological charge density, which undoubtedly integrates to an integer number.

At this point it is natural to introduce the notion of $\HP{1}$-projection. Namely,
one could replace the original potentials with corresponding $\sigma$-model induced gauge connection
\beq
\label{HP1:projection}
A_\mu ~ \to ~ A^\hp_\mu \equiv \bra{q}\diff_\mu\ket{q}\,,
\eeq
and then compare the physical content of the original and projected configurations.
It turns out that $A^\hp_\mu$ fields are extremely weak, the corresponding curvature
is by order of magnitude smaller than that of $A_\mu$ configurations. However, the detailed analysis
show that (we are necessarily brief here, see Refs.~\cite{Gubarev:2005rs,Boyko:2006jt} for details):
\begin{itemize}
\item projected theory is still confining, moreover, the projected string tension $\sigma_\hp$ exactly
reproduces the full one in the continuum limit;
\item topological aspects of the original fields are preserved. This could be verified
by various methods including modern fermionic ones.
\item investigation of the local observables, like topological charge or action densities,
suggests that only the leading  perturbative power divergences are
washed out by $\HP{1}$-projection.
In particular, the method allows to estimate rather precisely both the gluon condensate and its leading
(quadratic) power correction, the magnitudes of which are in full agreement with the existent
literature. Moreover, it was found that the topological fluctuations
behave quite non-trivially in the continuum forming the percolating three-dimensional domains
of sign coherent topological charge in the limit of vanishing lattice spacing.
The latter observation is again in full agreement with modern
literature~\cite{Horvath:2005cv,Horvath:2003yj,top:domains:A,top:domains:B}.
\end{itemize}

The overall likely conclusion is that the $\HP{1}$ embedding method is the perfectly gauge invariant
way to filter out only the leading UV divergent part from various observables leaving intact
the non-perturbative content of the original gauge configurations. However, there is a particular
disadvantage:
the  $\sigma$-model fields constructed this way are non-local in term of the original gauge potentials.
At present, there are two possibilities to estimate the range of
non-locality. First, the heavy quark potential extracted from $\HP{1}$-projected
Wilson loops (Figure 9 in the second paper of Ref.~\cite{Gubarev:2005rs})
appears to be concave at distances $|x| \lesssim 4$ (lattice units) for gauge configurations
thermalized at bare coupling $\beta = 2.60$.
Secondly, the $\HP{1}$-based topological density correlation function  $\mean{q_0 q_x}$
reveals a positive core at small distances, the size of which scales in physical
units~\cite{Boyko:2006jt}.
Both these observations indicate that the characteristic range of non-locality is (see Section~\ref{sec:numerics})
\beq
\label{HP1:non-local}
R_{\mathrm{non-local}}
\,\sqrt{\sigma_\hp} \lesssim 0.5\,.
\eeq
Note that although this number is parametrically large, Eq.\eq{HP1:non-local} provides only an upper bound.
We expect that for large enough Wilson loops the corrections due to the non-locality are negligible
provided that the linear extent and the width of the string is greater than $R_\mathrm{non-local}$.
For this reason in the numerical investigations below we systematically exclude data points for which
the relevant geometrical characteristics are smaller than $R_{\mathrm{non-local}}$.

\section{Observables and Theoretical Expectations}
\label{sec:observables}

Before going into details of the numerical simulations let us discuss the observables to
be measured. As far as the topological density $\mean{q^2}$
and $\HP{1}$-projected action density $\mean{s}$
are concerned, their construction is fully described in Refs.~\cite{Gubarev:2005rs,Boyko:2006jt}
and will not be repeated here. The prime objects of our study are the dimensionless quantities
\beq
\label{QC}
Q_\cC(x) = \frac{ \mean{ \, W(\cC) \, q^2(x) \, } }{ \mean{W(\cC)} \,  \mean{q^2} } - 1\,,
\eeq
\beq
\label{SC}
S_\cC(x) = \frac{ \mean{ \, W(\cC) \, s(x)\,} }{ \mean{W(\cC)} \, \mean{s} } - 1\,,
\eeq
which are the expectation values of squared topological density $\mean{q^2}$  and $\HP{1}$-projected
action density $\mean{s}$  in the presence of the external color sources represented  by the rectangular
$\cC = T \times R$ Wilson loop $W(\cC)$ in the fundamental representation. Note that $W(\cC)$ is to be evaluated
also on $\HP{1}$-projected potentials
\beq
W(\cC) = \tr \, \Pexp \Bigl\{ \int_\cC \, \dd t \, \dot{x}_\mu \, A^\hp_\mu \Bigr\}\,,
\eeq
where $t$ is natural parametrization of $\cC$ and dot denotes derivative with respect to $t$.
The normalization of both quantities $Q_\cC(x)$ and $S_\cC(x)$ is such that
in the case of negligible correlation with the confining string both quantities vanish.
Furthermore, suppression or enhancement of the topological/action density fluctuations
makes the corresponding quantity negative or positive, respectively.

It is amusing to note that the $\HP{1}$-projected Wilson loop has a pure geometrical interpretation. Indeed,
using Eqs.~(\ref{HP1:projector},\ref{HP1:projection}) one can show that
\beqn
\label{HP1:spin-factor}
& W(\cC) = \tr \, \Pexp \Bigl\{ \frac{i}{2}\int_\cC \, \dd t \, \sigma^{AB} \omega_{AB}(t) \Bigr\} = & \\
& = \Bigl[ \, \mathrm{det} \Bigl\{ \delta_{AB}\,\diff_t - \frac{1}{2}\omega_{AB} \Bigr\} \, \Bigr]^{1/2}\,, & \nonumber
\eeqn
where $\sigma^{AB} = \frac{i}{2} [ \gamma^A , \gamma^B ]$, $A,B = 1,...,5$ and
$\omega^{AB} = \frac{1}{2}( n^A \dot{n}^B - n^B \dot{n}^A) = n^{[A} \dot{n}^{B]}$ is the instantaneous
rotation speed of the color vector $n^A(t)$ naturally assigned to the contour $\cC$
(see, e.g., Ref.~\cite{spin-factor} for detailed derivation).
Therefore the $\HP{1}$-projected Wilson loop is nothing but the spin factor
associated with spinning particle propagating in fictitious five-dimensional color space, $n^A$
being the normalized tangent to its trajectory.
The spinning particle interpretation of the Wilson loop might have a deep consequences, however,
what precludes further analysis is the unknown measure of integration over $n^A$ field.
Indeed, at present we only know how to generate $n^A$ configurations numerically and have no idea
what is the key feature of the measure $\cD n$, which provides an area law for $\mean{W(\cC)}$.
In principle, one could speculate that the measure $\cD n$ might be taken standard and
the area law would still be captured provided that the flat metric $\delta_{AB}$
in \eq{HP1:spin-factor} is replaced by some effective one $g_{AB}$ describing geometry
of ambient 5-dimensional space.
This way of thinking resembles somewhat the AdS/QCD framework (see, e.g., Ref.~\cite{AdS}),
however, it would lead us too away from the purposes of present discussion.

\begin{figure}[t]
\centerline{\psfig{file=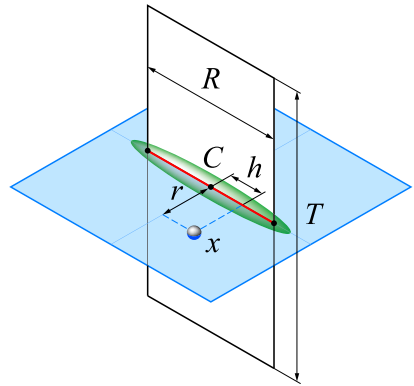,width=0.3\textwidth,silent=}}
\caption{(Color online) The geometry of our simulations. The topological/action densities are
calculated at the point $(r,h)$, where $r$ and $h$
are, respectively, the transverse and longitudinal coordinates with respect to the geometrical center $C$
of the rectangular $T \times R$ Wilson loop.}
\label{fig:geometry}
\end{figure}

The geometry relevant to the correlation functions \eq{QC}, \eq{SC} is shown in Figure~\ref{fig:geometry}.
For the rectangular $T \times R$ Wilson loop the chromoelectric string appears
between the static quark-antiquark pair at the moment $t=0$.
Once created, the string develops in full at the time $t=T/2$
when the numerical evaluation of the string profile is performed.
Finally, the string disappears at the time $t=T$ when the quark and the antiquark are annihilated.
For large enough $T$ the problem evidently possesses spacial cylindrical symmetry,
which suggests to use coordinates $x = (r, h)$, where $r$ and $h$ are, respectively,
the transverse and longitudinal distances to the geometrical string center $C$.

As is well known, at the quantum-mechanical level the string must fluctuate around classical
(minimum action) configuration.
L\"uscher~\cite{Luscher:1980iy} has obtained long ago
a quantitative estimation of the quantum fluctuations of a bosonic string.
According to Ref.~\cite{Luscher:1980iy} one generally expects that the deviation
of the string position with respect to its classical limit should
follow a Gaussian distribution with a calculable width.
The quantum fluctuations of the string should also affect distributions of all other quantities
associated with the string.
In particular, the profiles of the action and topological densities should become
noticeably wider due to the string fluctuations compared to the that for the classical
(non-fluctuating) string.
Therefore, one expects generically that the transverse slices of the correlation functions
\eq{QC}, \eq{SC} at the geometrical center $h=0$ of the string should follow the L\"uscher scaling law:
\beq
X_\cC(r,R) = - C_X(R) \, \exp\Bigl\{ - \frac{r^2}{\delta^2_X(R)} \Bigr\}\,,
\label{eq:gauss}
\eeq
where $X$ stands for either the topological charge or the action, $X = Q, S$.
This distribution is characterized by two quantities which are the amplitude
$C_X$ and the width $\delta_X$ of the transverse string fluctuations.
Note that both these quantities depend in general on the length of the string $R$.
In particular, the squared width of the bosonic string is known~\cite{Luscher:1980iy}
to increase logarithmically for sufficiently long strings:
\beqn
\delta^2_X(R) = \frac{\zeta_X}{\pi \sigma} \ln \frac{R}{R^\mathrm{core}_X}\,.
\label{eq:delta:th}
\eeqn
Here $\sigma$ is the tension of the string and $\zeta_X$ is a coefficient of proportionality
which is expected to be unity in $(3+1)$ dimensions
\beqn
\zeta_X = 1\,.
\label{eq:C:theor}
\eeqn
Note that the range of qualitative applicability of the result~\eq{eq:gauss} is dictated by purely geometrical
reasons: the length $R$ of the string should be larger than some critical value $R^{\mathrm{core}}_X$
of order of the string width.
In the case of unprojected action density, which is related to the heavy quark potential
via the action sum rules~\cite{ref:sum-rules},
one assumes that the coefficient $\sigma$ in Eq.~(\ref{eq:delta:th}) equals to the physical string tension.
For the $\HP{1}$-projected action density we also expect the same:
although the sum rules are not applicable any longer,
physics-wise it seems natural that $\sigma$ must be equal to the $\HP{1}$-projected string tension.
However, for the topological density the arguments break down and
the scale of the width of the topological fluctuations in the vicinity of the string
should be examined in the numerical
simulations.

\section{Details of Numerical Simulations}
\label{sec:numerics}

\begin{table}[t]
\centerline{\begin{tabular}{|c|c|c|c|c|c|} \hline
$\beta$ & \rule{0pt}{11pt}
$a\sqrt{\sigma_{\mathrm{full}}}$ & $a \sqrt{\sigma_\hp}$ & $N^{\mathrm{conf}}$ & $V^{\mathrm{lat}}$ & $V^{\mathrm{phys}}
\cdot \sigma^2_\hp$ \\ \hline
2.510 & 0.176(2) & 0.119(3) & 60 & $32^4$ & $[3.8(1)]^4$ \rule{0pt}{11pt} \\
2.600 & 0.131(2) & 0.105(2) & 70 & $28^4$ & $[2.9(1)]^4$ \\ \hline
\end{tabular}}
\caption{The parameters of our simulations.}
\label{tab:params}
\end{table}

Our measurements were performed on two sets (Table~\ref{tab:params})
of statistically independent $SU(2)$ gauge configurations generated with standard Wilson action.
These thermalized gauge ensembles were used only to produce the embedded $\HP{1}$ $\sigma$-model configurations,
the procedure being exactly the same as in the second paper of Ref.~\cite{Gubarev:2005rs}.
The physical scale for the obtained $\sigma$-model data sets could be fixed by the requirement
that either the full string tension $\sigma_{\mathrm{full}}$
or its $\HP{1}$-projected counterpart $\sigma_\hp$
equals to the conventional value $(440~\mathrm{MeV})^2$.
Physically these two prescriptions are equivalent since it is known that the $\HP{1}$-projected
string tension coincides with the full one in the continuum limit.
However, at necessarily finite $\beta$ values
a particular choice might be preferable for scaling checks.
For us the prescription  $\sqrt{\sigma_{\mathrm{full}}} = 440~\mathrm{MeV}$
seems to be unnatural
since after all the original gauge configurations were not used in the measurements.
What is physically relevant in the present context is the $\HP{1}$-projected string tension
and we set the scale by fixing $\sigma_\hp$ to the above value.
Note that since this prescription differs from the conventional one
we use the dimensionless combinations like $R\sqrt{\sigma_\hp}$ instead of the usual fermi units.

The string profiles may obviously be affected by the finiteness of the temporal extension $T$.
To reduce these finite-time effects we used the conventional APE-smearing procedure~\cite{ref:smearing}
for spatial pieces of the trajectory $\cC$ and checked independence of our results against variation of $T$.
The geometrical setup described above makes it convenient to consider Wilson loops of only even
temporal and spacial extents  in lattice units. Furthermore,  the non-locality issue, Eq.~(\ref{HP1:non-local}),
forces us to disregard the smallest loops. Thus we confined ourselves to planar $T\times R$ Wilson
contours with $T, R = 4, ..., 14$ with understanding that $T,R = 4, 6$ data points must be used with care.

\section{Topological Density in the String}
\label{sec:topology}

\begin{figure}[t]
\includegraphics[scale=0.65,clip=false]{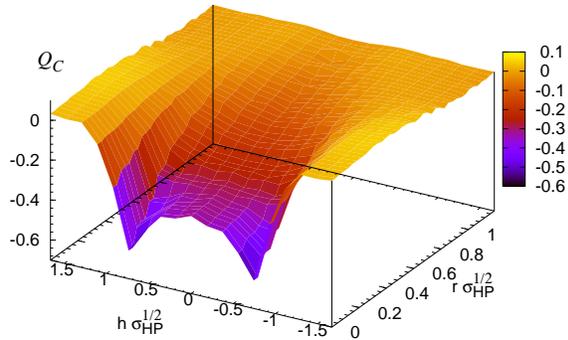}\\[3mm]
\caption{(Color online) Correlation function \eq{QC} at quark-antiquark separation $R \sqrt{\sigma_\hp} = 1.47(3)$
measured on $\beta=2.60$ data set. Temporal extent of the Wilson loop is $T \sqrt{\sigma_\hp} = 1.05(2)$.}
\label{fig:profile:3D}
\end{figure}

The qualitative topological portrait of the confining string is revealed by the correlation function \eq{QC},
shown in Figure~\ref{fig:profile:3D} for our $\beta=2.60$ data set.
As is apparent from  this  Figure the quantity $Q_\cC$
is always negative in the vicinity of the string while approaching zero far away from quark-antiquark pair.
Thus the confining flux tube suppresses the topological fluctuations.
Note also that the suppression is maximal at the positions of quarks, which
are sitting at the absolute minima of the topological density profile
(this observation is in agreement with Ref.~\cite{string:topcharge:first,string:topcharge}).
Moreover, the quarks look like the sources of Coulomb potential in the topological density,
which by itself a rather striking result especially in view of the corresponding action density
profile to be discussed below.

\begin{figure}[t]
\vskip -7mm
\includegraphics[scale=0.65,clip=false]{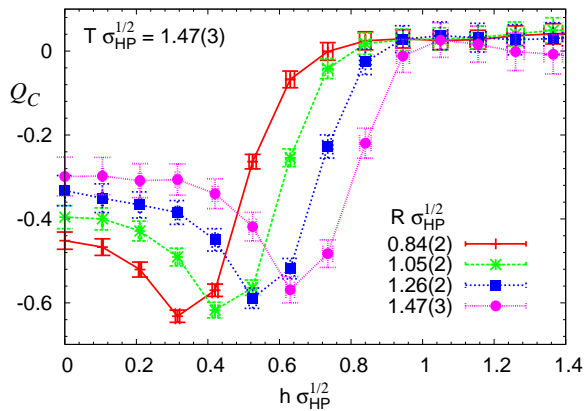}\\[3mm]
\caption{(Color online) The longitudinal slices ($r=0$) of the topological density correlation function \eq{QC}
at various quark-antiquark separations $R$ at $\beta=2.60$.
Only the positive half ($h \geqslant 0$) of the axis is shown, lines are plotted to guide the eye.}
\label{fig:profile:longitudinal}
\end{figure}

The typical longitudinal ($r=0$) slices of the topological density correlation function \eq{QC}
are plotted in Figure~\ref{fig:profile:longitudinal} for various lengths $R$ of the confining string.
The minima of the topological density are reached in the vicinities of the quark and anti-quark,
$h_{\min} \approx R/2$. As one goes outward the string, $h> h_{\min}$, the quantity
$Q_\cC$ rapidly approaches zero indicating that topological charge density quickly comes to its
vacuum expectation value outside the string.
To the contrary, if one moves toward the string center, $h \to 0$, the magnitude of $Q_\cC$ remains
negative slowly approaching its limiting value at $h=0$.

\begin{figure}[t]
\includegraphics[scale=0.65,clip=false]{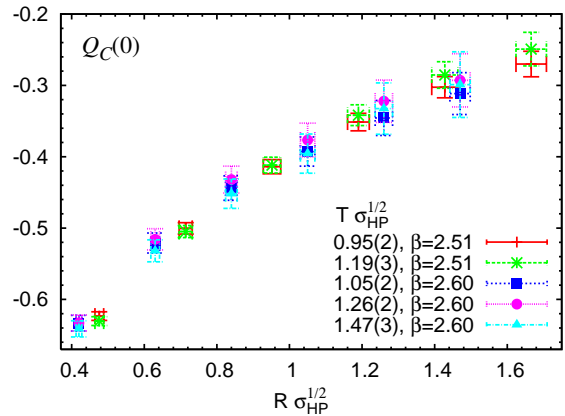}\\[3mm]
\caption{(Color online) The topological density correlation function~\eq{QC}
at the geometrical center of the string ($h=r=0$) versus the quark separation $R$
for the Wilson loops of various temporal extension.}
\label{fig:center:point}
\end{figure}

The degree of the suppression of the topological density at the geometrical center of the string
$r = h = 0$ varies with the string length. This property is visualized for both our data sets
and for various temporal extensions $T$ of the Wilson loops on Figure~\ref{fig:center:point}.
One can see that the topological density at the string center tend to its
vacuum value as the string gets longer. This feature is in agreement with
the picture of the fluctuating string:
as the string gets longer its transverse fluctuations increase in amplitude and the string
gets more dispersed in the space.
As the consequence, the correlation of all quantities (including the topological charge density)
with the string gets weaker  at the geometrical center  with increasing string length.

\begin{figure}[t]
\includegraphics[scale=0.65,clip=false]{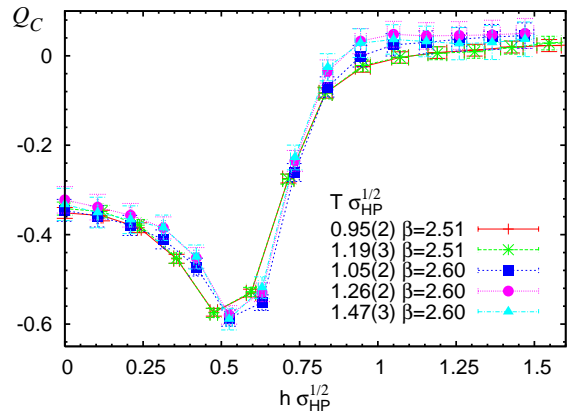}\\[3mm]
\caption{(Color online) Example of the longitudinal profiles at $r=0$ of the correlator \eq{QC}
at string lengths $R\sqrt{\sigma_\hp} = 1.19(3)$ for $\beta=2.51$ and
$R\sqrt{\sigma_\hp} = 1.26(2)$ for $\beta=2.60$
and various time extensions of the Wilson loop.}
\label{fig:profile:time}
\end{figure}

Another information imprinted in Figure~\ref{fig:center:point} is small sensitivity
of the central value of the topological density correlator \eq{QC} against the variation of
the temporal extent of the Wilson loop $T$.
The typical dependence of the string profile for $h > 0$ on the time extension of the Wilson loop
is presented on Figure~\ref{fig:profile:time} for the quark-antiquark separations $R\sqrt{\sigma_\hp} = 1.19(3)$
(data set $\beta=2.51$) and $R\sqrt{\sigma_\hp} = 1.26(2)$ ($\beta=2.60$).
It is seen that the string profile is indeed robust against the time passed between creation
and annihilation of the quark pair.
Therefore we conclude that within our accuracy
the dependence of the results on the time extension $T$ is negligible.

\begin{figure}[t]
\includegraphics[scale=0.65,clip=false]{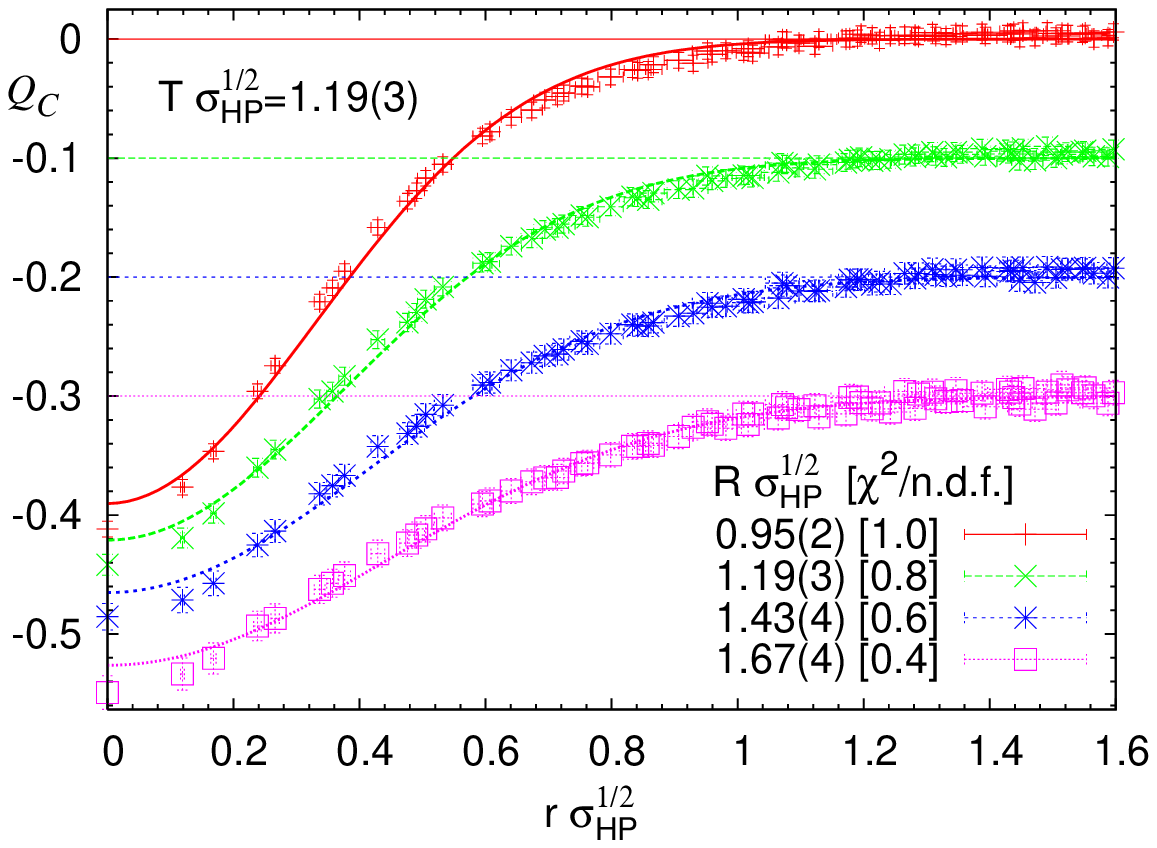}\\
\includegraphics[scale=0.65,clip=false]{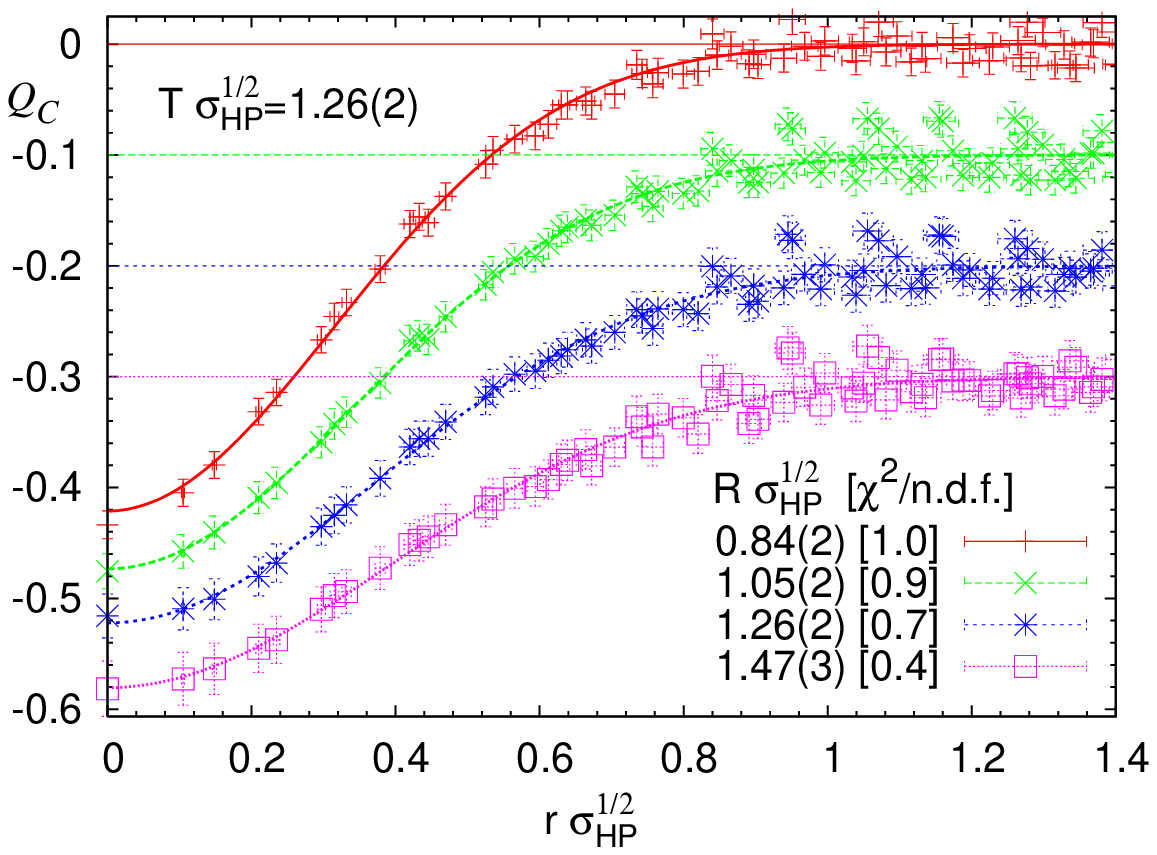}\\[3mm]
\caption{(Color online) The distributions of the topological charge fluctuations~\eq{QC}
in the transverse direction $r$ around the geometrical string center $h = 0$.
For the sake of clarity the data points were $y$-shifted, and
the corresponding $Q_\cC=0$ levels are shown by thin horizontal lines.
Upper and lower panels refer, respectively, to the
$\beta=2.51$ and $\beta=2.60$ data sets.
Best fits by the Gaussian function~\eq{eq:gauss} are shown by thick lines.
The string lengths $R$, the temporal extensions $T$ and the reduced $\chi^2$ values are indicated as well.}
\label{fig:Q2:fits}
\end{figure}

\begin{figure}[t]
\includegraphics[scale=0.65,clip=false]{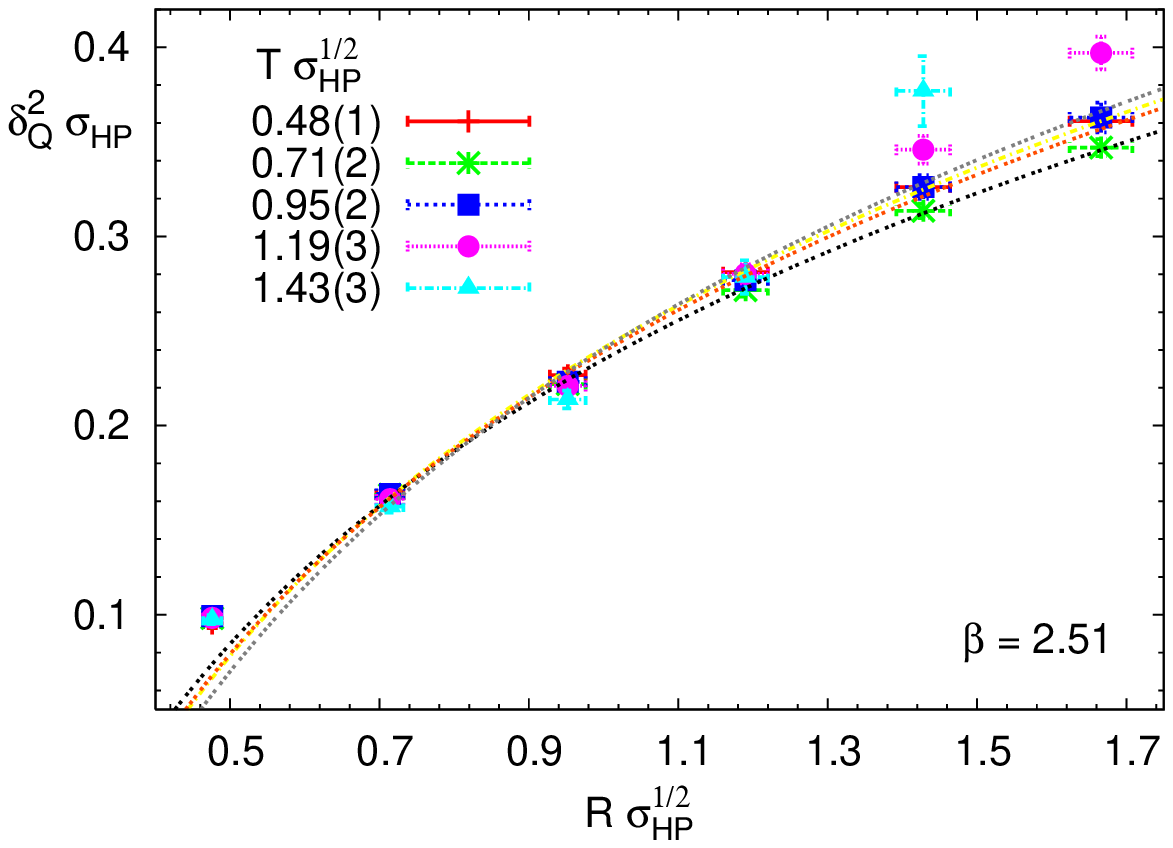}\\
\includegraphics[scale=0.65,clip=false]{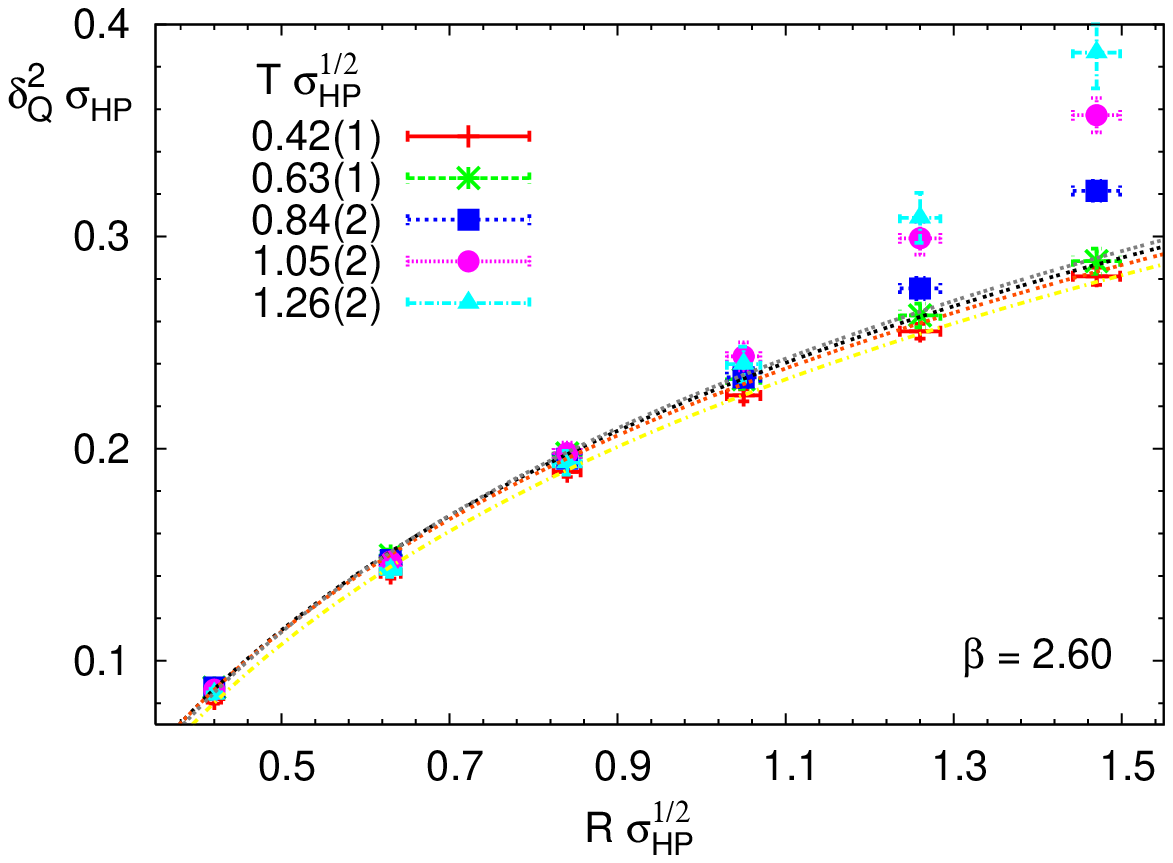}\\[3mm]
\caption{(Color online) The squared string width $\delta^2_Q$ {\it vs.} the string length $R$ for
various time extensions of the Wilson contour. Data points had been obtained with
$\beta=2.51$ (top) and $\beta=2.60$ (bottom) sets.  The lines represent the best fits
by the function \eq{eq:delta:th} (the details are given in the text).}
\label{fig:luscher:fits}
\end{figure}

In order to check the validity of the L\"uscher formula~\eq{eq:gauss}
for the width of the topological charge fluctuations around the string
we fitted the transverse slice of the correlation function \eq{QC} at the string center $h=0$
by the Gaussian ansatz~\eq{eq:gauss}.
We find that the scaling law~\eq{eq:gauss} well describes both our data sets as one can see
from various examples of the fits shown in Figure~\ref{fig:Q2:fits}.
The only notable deviations from \eq{eq:gauss} was observed for small string lengths,
$R\sqrt{\sigma_\hp} \lesssim 0.7$, which, however, is to be expected  since
Eq.~(\ref{eq:gauss}) is valid only for large $R$.
Therefore in terms of the topological density the string formation length,
at which the asymptotic behavior \eq{eq:gauss} settles in,
could be estimated as $R\sqrt{\sigma_\hp} \approx 0.7$ and is in agreement
both with the literature~\cite{Bali:1994de} and with the corresponding result coming
from the action density profiles (see below).
In turn, the Gaussian fits allow to obtain the corresponding parameters.
On the bottom panel of Figure~\ref{fig:luscher:fits} we plot the squared string width $\delta^2_Q(R)$
obtained for $\beta=2.60$ data set as a function of the string length $R$ for various
temporal extensions $T$ of the Wilson contour.
One obvious thing to note here is that $\delta^2_Q(R)$ data points
are indeed compatible with logarithmic dependence \eq{eq:delta:th} in the whole $R$ interval for small
$T \sqrt{\sigma_\hp} = 0.42(1), 0.63(1)$, while for  $T \sqrt{\sigma_\hp} \gtrsim 0.84(2)$
they start to deviate from (\ref{eq:delta:th}) for large string lengths.
In fact, for $T \sqrt{\sigma_\hp} = 0.84(2), 1.05(2)$ the total set of $R$ data points corresponds
rather to the linear dependence of $\delta^2_Q$ on $R$,
which is presumably a manifestation of finite size effects. Indeed, the considerations~\cite{Bali:1994de} showing that
finite size corrections are negligible even for largest loops are not applicable in our case:
for $\HP{1}$ $\sigma$-model induced gauge fields the relevant center symmetry transformations do not exist.
Therefore the finite volume effects are expected to be much more severe than they are in usual approaches and indeed
at largest available $T \sqrt{\sigma_\hp} = 1.26(2)$ the squared string width increases
with $R$ even more rapidly showing somewhat irregular behavior for large quark-antiquark separations.
We attribute the deviation of the string width from logarithmic law to the boundary conditions in the temporal
and spatial directions: the long enough string interacts with its temporal and spatial replicas.
The strength of this artificial interaction is rising with increase of the area of the string worldsheet
achieved at large $R$ and $T$.
The assumption of strong finite volume effects present for $\beta=2.60$ configurations
is justified below (see also Section~\ref{sec:action}).  However, for small $T$ the finite volume
corrections are surely negligible, which permits us to fit $\delta^2_Q(R)$ to Eq.~(\ref{eq:delta:th})
in the range $R \sqrt{\sigma_\hp} \gtrsim 0.7$ for $T \sqrt{\sigma_\hp} = 0.42(1), 0.63(1)$
and in the interval $0.7 \lesssim R \sqrt{\sigma_\hp} < 1.2$ at $T \sqrt{\sigma_\hp} = 0.84(2), 1.05(2)$.
The lower bound on $R$ is due to Eq.~(\ref{HP1:non-local}) and because
the Gaussian fit \eq{eq:gauss} is not adequate at small distances resulting in rather large
$\chi^2$ values.  The outcome of these fits is used below.

In order to check finite volume effects we performed the same calculations on our $\beta=2.51$ configurations,
for which the physical volume is much larger. The results are presented on the top panel of Figure~\ref{fig:luscher:fits},
from which it is apparent that the data points are much more stable in this case.
Namely, the dependence $\delta^2_Q(R)$ follows the logarithmic law \eq{eq:delta:th}
up to the time scale of order $T \sqrt{\sigma_\hp} \lesssim 1.2$.
However, at even larger $T$ we again see notable deviations from Eq.~(\ref{eq:delta:th})
the source of which is not so evident for us at present.
Most probably, the errors bars are underestimated for the  corresponding largest Wilson loops
and much more statistics in needed to confirm \eq{eq:delta:th} at
$T \sqrt{\sigma_\hp} \gtrsim 1.2$, $R \sqrt{\sigma_\hp} \gtrsim 1.4$.

To summarize, our numerical data on the string width obtained from the topological density correlation function
\eq{QC} indeed favors the logarithmic widening \eq{eq:delta:th} of the confining string.
Unfortunately, $\beta=2.60$ data set has relatively small physical volume which disturbs
the expected dependence $\delta^2_Q(R)$.  Anyway,
for not so large time extent of the Wilson loops logarithmic string broadening is clearly seen.
As far as the measurements at $\beta=2.51$ are concerned, the corresponding physical volume is much larger
and the logarithmic widening is confirmed for much larger Wilson loops.

\begin{figure}[t]
\includegraphics[scale=0.65,clip=false]{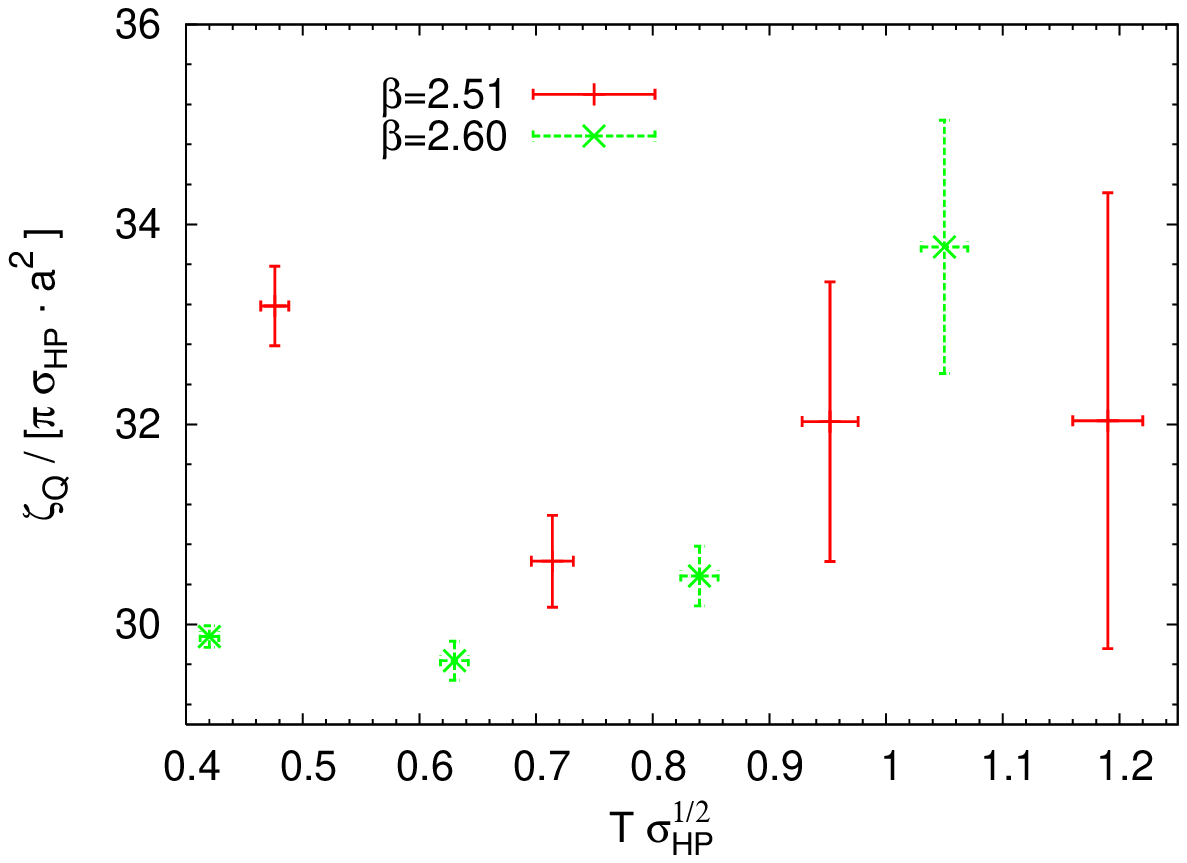}\\
\includegraphics[scale=0.65,clip=false]{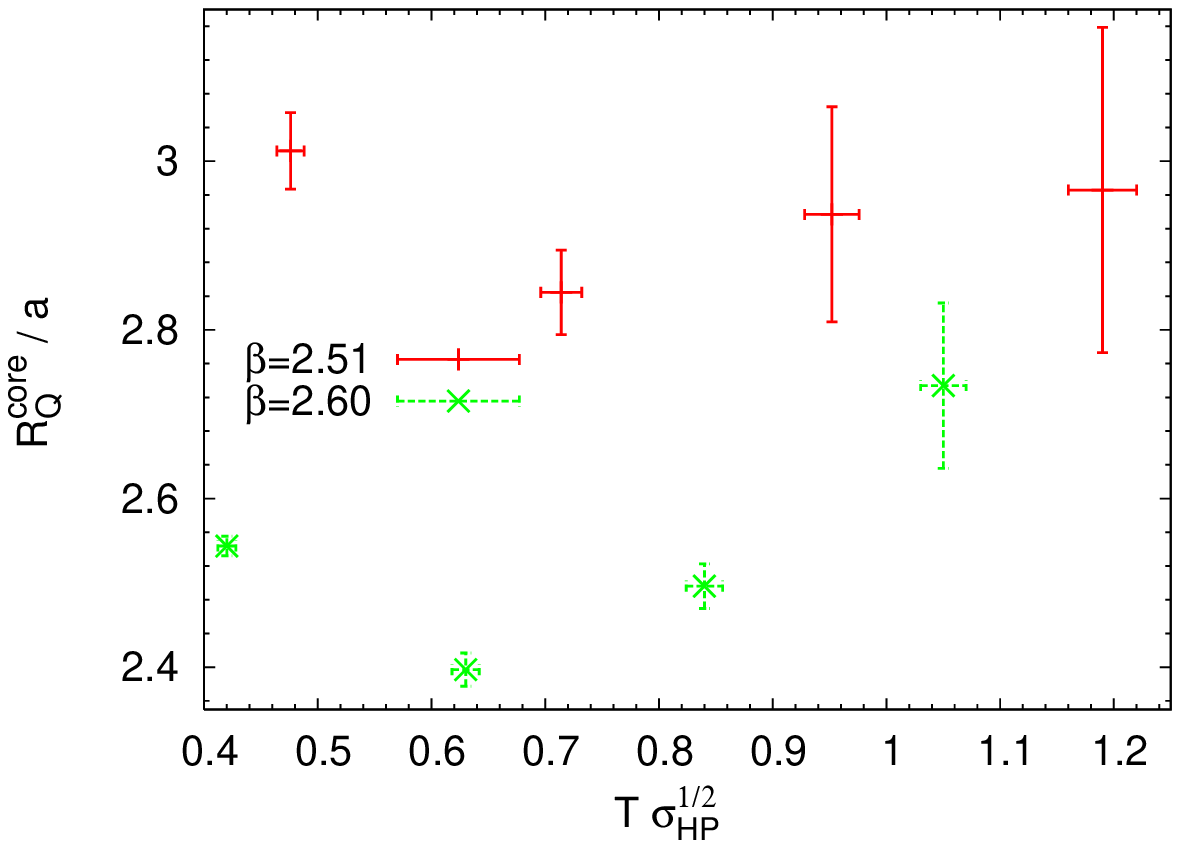}\\[3mm]
\caption{(Color online) The coefficients $\zeta_Q/[\pi \sigma_\hp \cdot a^2]$ (top) and $R^{\mathrm{core}}_Q/a$
(bottom) of the L\"uscher scaling law~\eq{eq:delta:th} as determined from Wilson loops
of various time extension $T$  for both our data sets. Note that
all quantities are given in the lattice units.}
\label{fig:bestfit:t}
\end{figure}

The fits of the Gaussian string profiles~\eq{eq:gauss} allow us to estimate the  dimensionless parameters
$\zeta_Q$ and $R^{\mathrm{core}}_Q$ of the L\"uscher scaling law~\eq{eq:delta:th}.
According to the theoretical expectation the parameter $\zeta_Q$ must of order one
provided that the width $\delta_Q$ is given in terms of the physical string tension, while
there are essentially no theoretical predictions for the string core width $R^{\mathrm{core}}_Q$.
However, the relevant question for both quantities is whether
they are given in physical units or are of order lattice spacing.
In the upper and lower panels of Figure~\ref{fig:bestfit:t} we plot
the parameters $\zeta_Q/[\pi\sigma_\hp \cdot a^2]$ and $R^{\mathrm{core}}_Q/a$, respectively,
obtained by the logarithmic fits~\eq{eq:delta:th} as a functions of the temporal string extension $T$.
It is apparent that $T$-dependence is indeed mild in both cases. At the same time,
it is essential that the parameters are expressed in the units of UV cutoff,
which we found the only possibility to make $\beta=2.51$ and $\beta=2.60$ data points qualitatively
consistent. Hence the string width and the size of the string core,
as they are seen by the topological fluctuations, are likely to be at the scale of the lattice spacing.
Unfortunately, this result is mainly qualitative
and is not fairly convincing because we consider only two lattice spacings while
the logarithmic fits are weakly sensitive to variations of $R^{\mathrm{core}}_Q$.
For these reasons we refrain from definite conclusions since more data is needed to clarify the issue.

\section{Action Density in the String}
\label{sec:action}

\begin{figure}[t]
\includegraphics[scale=0.65,clip=false]{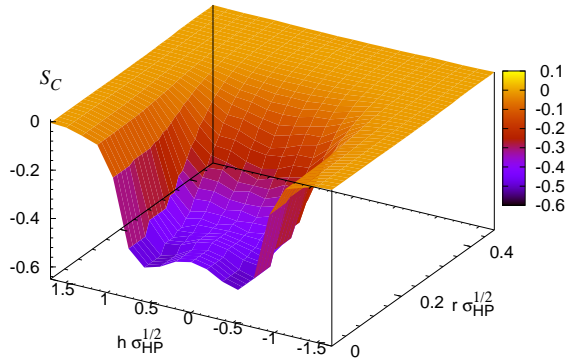}\\[3mm]
\caption{(Color online) The action density correlation function \eq{SC} measured
on $\beta=2.51$ configurations for the string length $R\sqrt{\sigma_\hp} = 1.67(4)$
and for  the temporal extension of the Wilson loop $T\sqrt{\sigma_\hp} = 1.19(3)$.}
\label{fig:action:profile:3D}
\end{figure}

In this Section we repeat the above investigations of the string profiles
using the $\HP{1}$-projected action density
\beq
s = \tr F^\hp_{\mu\nu} F^\hp_{\mu\nu}\,,
\eeq
where the inessential normalization factor was omitted.
Note that in this case the obtained statistics is much better allowing more precise analysis.
The typical distribution of the action density around the string is presented on
Figure~\ref{fig:action:profile:3D}.
It is clear from the figure that the action density is indeed suppressed in the vicinity
of the string in agreement with the analogous result for unprojected fields~\cite{ref:sum-rules}.
At the same time, there are a few noticeable differences between the action density profile
of the string  and the corresponding topological charge correlation function, Figure~\ref{fig:profile:3D}.
First, the positions of the test quark and the anti-quark are explicitly noticeable in the string topological
profile as these positions correspond to the two global minima of $Q_\cC$.
However, the location the test charges in the action density correlator
is not seen clearly. Second, the transverse distribution of the action is seemingly wider
compared to the one of the topological charge. Finally, the data for the action comes with much less
noise compared to one of the topological distribution at the same numerical statistics.

\begin{figure}[t]
\includegraphics[scale=0.65,clip=false]{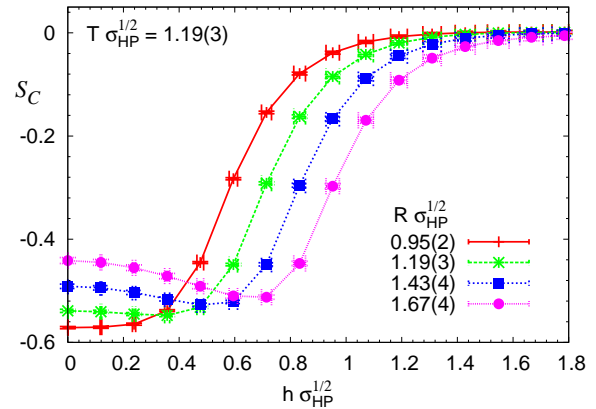}\\[3mm]
\caption{(Color online) The longitudinal profile ($r=0$) of the action density correlator $S_\cC$ obtained with
$\beta=2.51$ data set. The positive half ($h \geqslant 0$) of the axis is shown for various
indicated string lengths $R$. The lines are plotted to guide eye.}
\label{fig:action:profile:longitudinal}
\end{figure}

\begin{figure}[t]
\includegraphics[scale=0.65,clip=false]{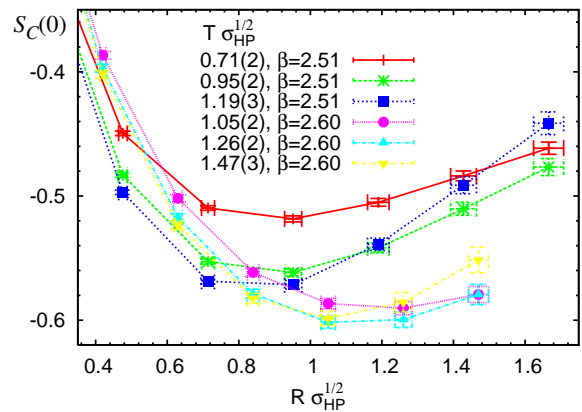}\\[3mm]
\caption{(Color online) The action density~\eq{QC} at the center of the string ($h{=}r{=}0$) {\it vs.}
the string length $R$ at various time extensions~$T$.
}
\label{fig:action:center:point}
\end{figure}

In order to illustrate the above  observations we plot in Figure~\ref{fig:action:profile:longitudinal}
the longitudinal profiles of the action density and compare them with the same quantity for the topological
charge plotted in Figure~\ref{fig:profile:longitudinal}.
One notices that the global minimum of the quantity $S_\cC$ shifts with increasing distance between test
color charges, being located at the string center point for relatively small $R$. As $R$ enlarges the minimum
shifts roughly to the string end-point, however, it remains much wider compared to that of the topological density.
This property is also illustrated in Figure~\ref{fig:action:center:point} in which the correlation function
\eq{QC} at the geometrical center $r=h=0$ of the string is plotted as a function of the string length $R$.
We conclude qualitatively that the location of the quarks becomes visible in the  longitudinal action
density profile at string length about $R \sqrt{\sigma_\hp} \approx 1.0$. Note that the apparent discrepancy
in this number between $\beta=2.51$ and $\beta=2.60$ data points is presumably due to the finite volume effects
discussed in previous section.

\begin{figure}[t]
\includegraphics[scale=0.65,clip=false]{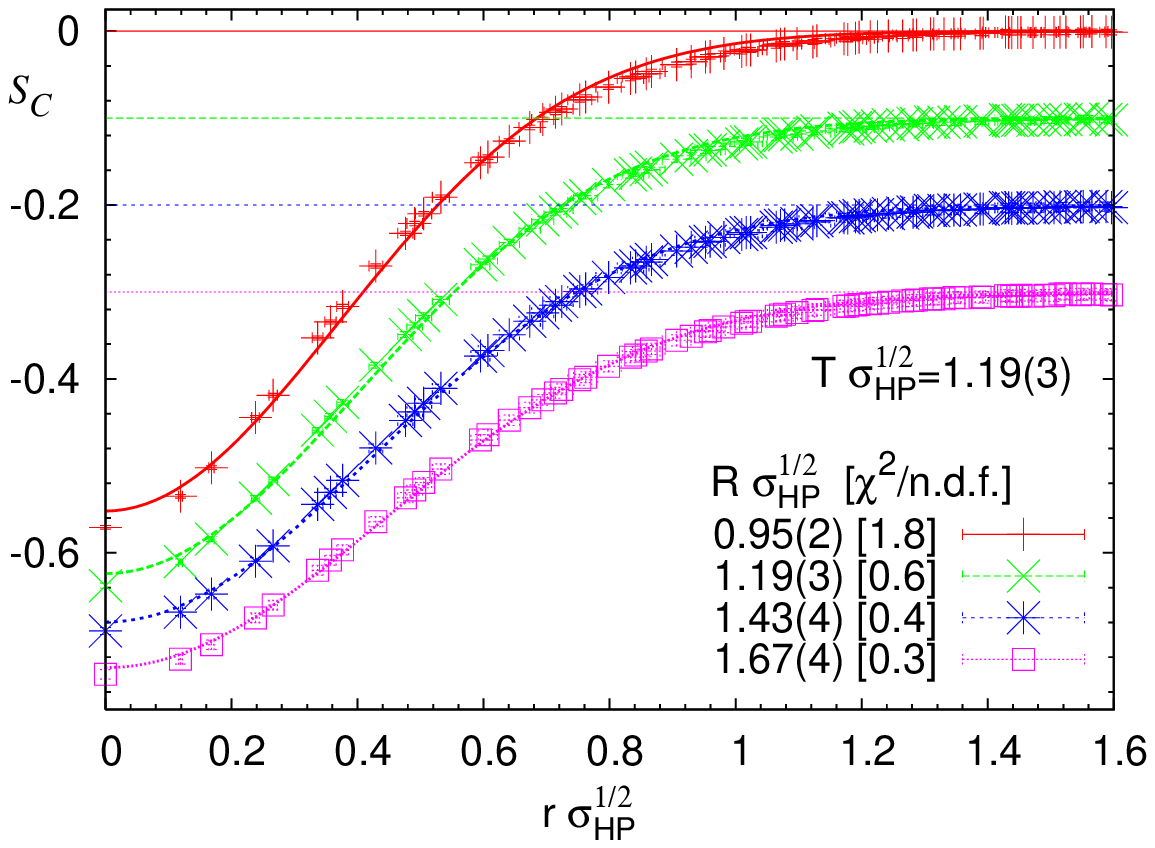}\\
\includegraphics[scale=0.65,clip=false]{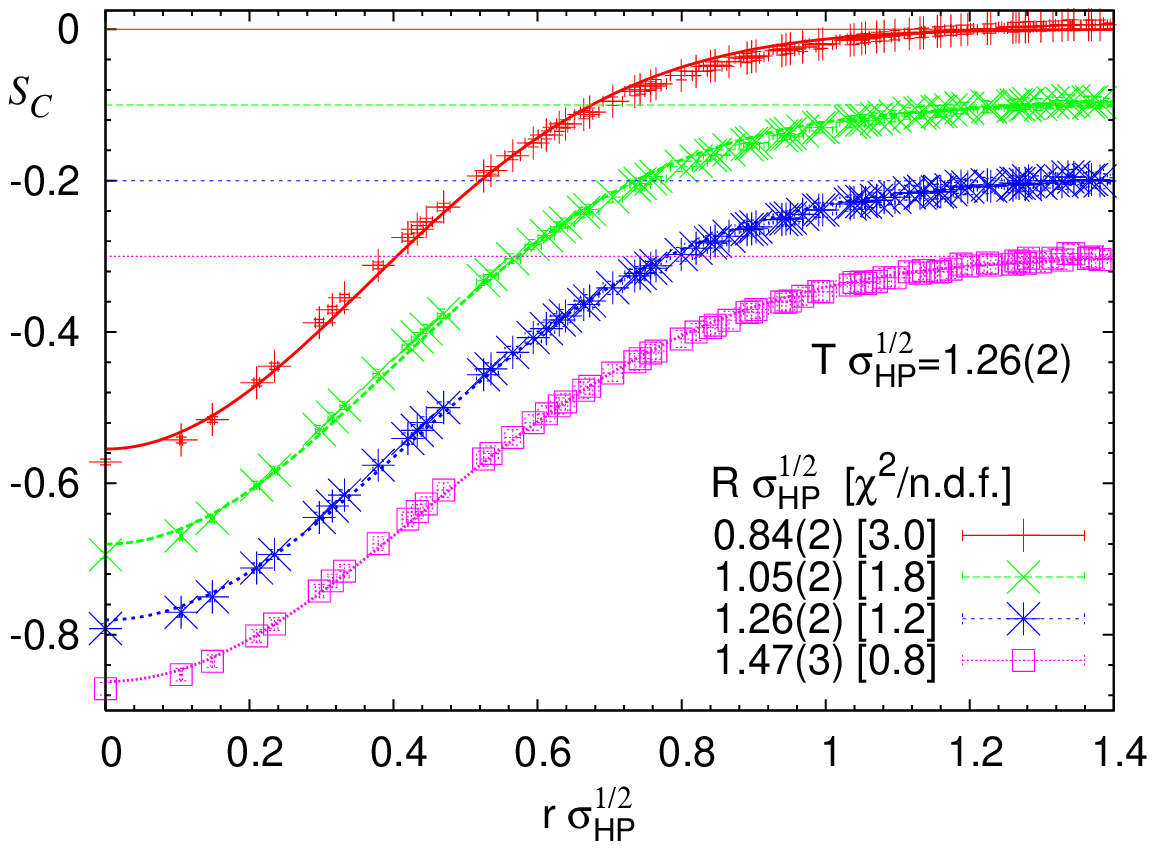}\\[3mm]
\caption{(Color online) Transverse slices of the action density correlation function \eq{SC}
at the geometrical center of the string $h=0$ for various temporal $T$ and spacial $R$ extensions
of the Wilson contour. Upper (lower) panel refers to $\beta=2.51$ ($\beta=2.60$)
data set. For clarity reasons data points were  $y$-shifted by $-0.1$, $-0.2$ and $-0.3$
(the corresponding $S_\cC=0$ levels are shown by thin horizontal lines).
The fits of the distributions by the Gaussian function~\eq{eq:gauss} are shown by thick lines.}
\label{fig:action:fits}
\end{figure}

As it was done in the case of the topological charge, we study the transverse slices of the action density
correlation function \eq{SC} at the geometrical center $h=0$ of the string.
Generically, one expects that the Gaussian distribution \eq{eq:gauss} should adequately describe the data
provided that the string length is large enough.
It turns out that indeed the transverse action density distribution around the string
obeys \eq{eq:gauss} for $R \sqrt{\sigma_\hp} \gtrsim 0.7$ as is illustrated on Figure~\ref{fig:action:fits}.
For smaller string lengths the data shows significant deviations from the Gaussian
ansatz indicating that the string formation length is not reached yet.
Therefore, the string width could be extracted reliably from fits only for
$R \sqrt{\sigma_\hp} > 0.7$ and in the considerations below we restrict ourselves to these distances.

\begin{figure}[t]
\includegraphics[scale=0.65,clip=false]{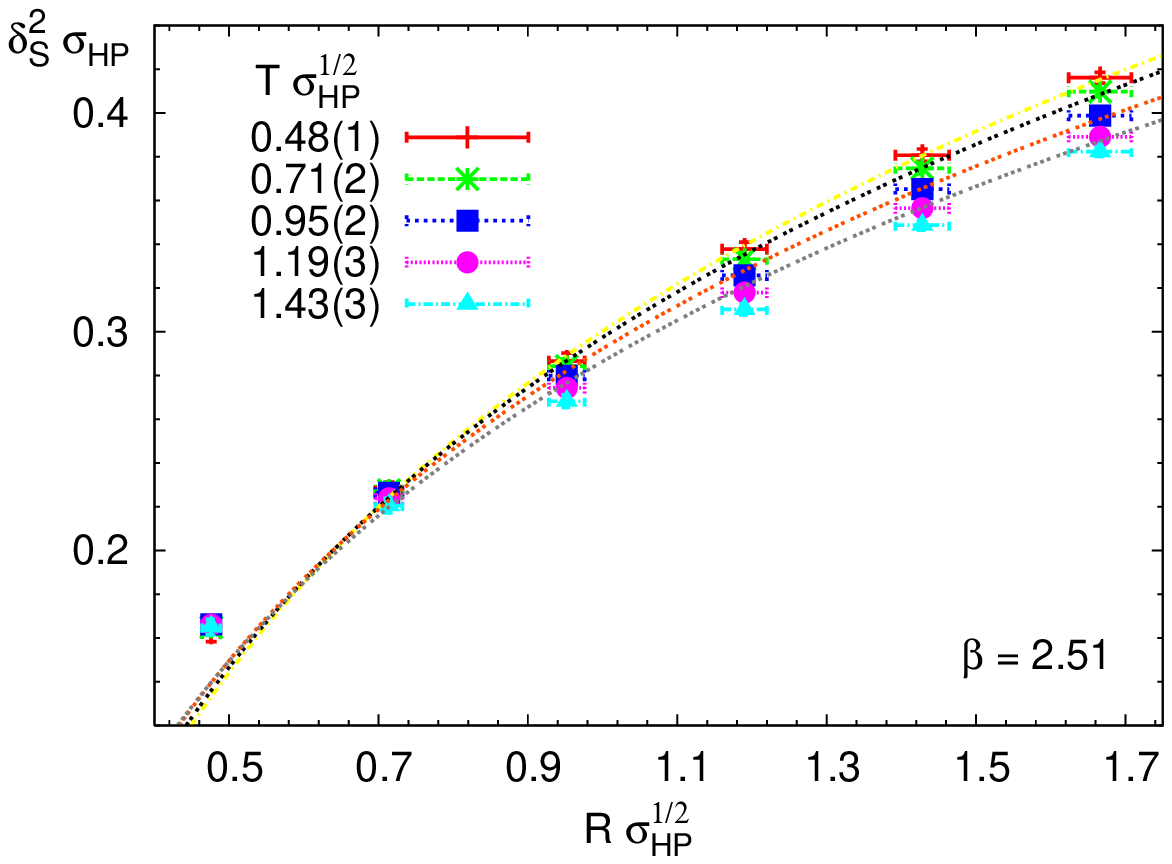} \\
\includegraphics[scale=0.65,clip=false]{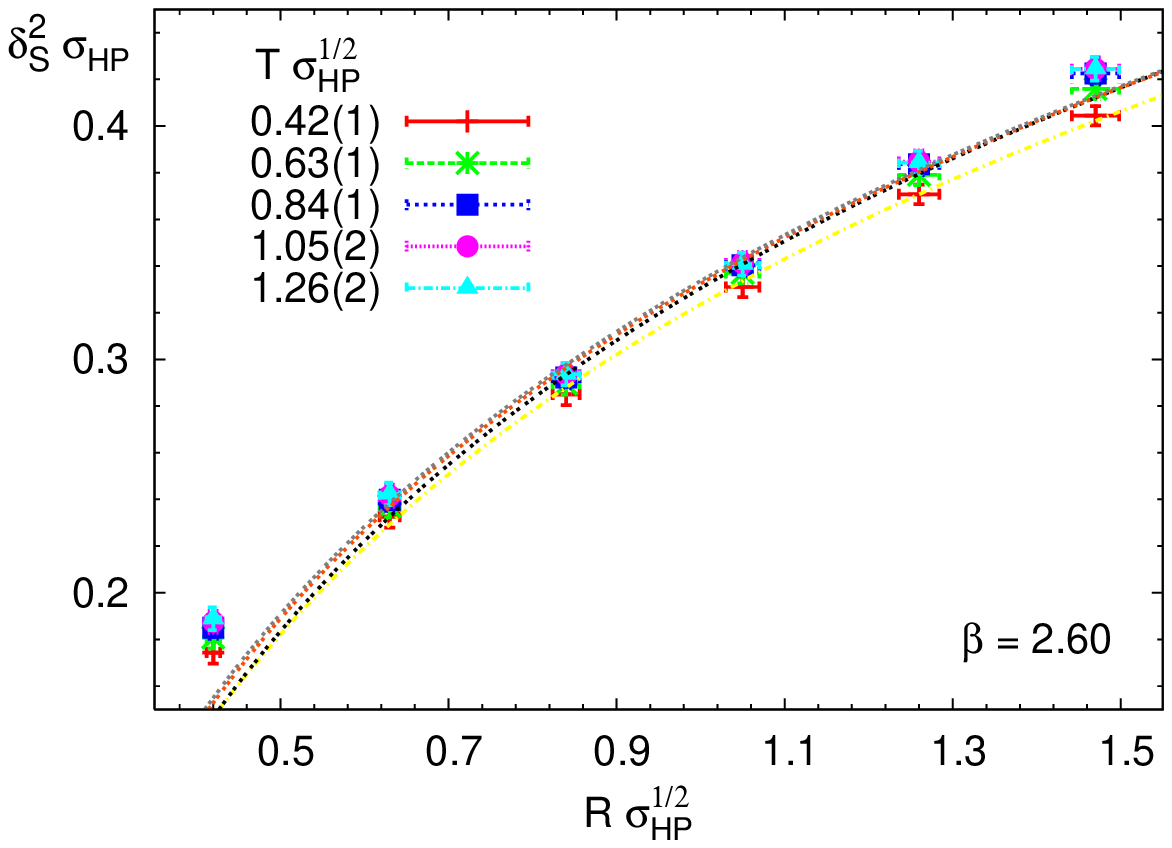} \\[3mm]
\caption{(Color online) The squared string width $\delta^2_S$ {\it vs.} the string length $R$ for
various time extensions  $T$ of the Wilson contour. Data points corresponds to
$\beta=2.51$ (top) and $\beta=2.60$ (bottom) sets. The lines represent the best fits to
\eq{eq:delta:th}.}
\label{fig:action:delta}
\end{figure}

As far as the parameters of the Gaussian distribution are concerned, we plot the corresponding
squared width of the action density profile obtained at various $R$ and $T$ on Figure~\ref{fig:action:delta}.
As is apparent from that Figure, the data points for $\delta^2_S(R)$ at each $\beta$-value
confirm the logarithmic string widening for all available $T$ in the entire range
of quark-antiquark separations beyond the string formation length.
Moreover, the function $\delta^2_S(R)$ appears to depend very mildly on $T$. However, basing on our
experience with the topological density distribution we excluded from our analysis
the largest $T$ data points, which presumably suffer from finite volume corrections.

\begin{figure}[t]
\includegraphics[scale=0.65,clip=false]{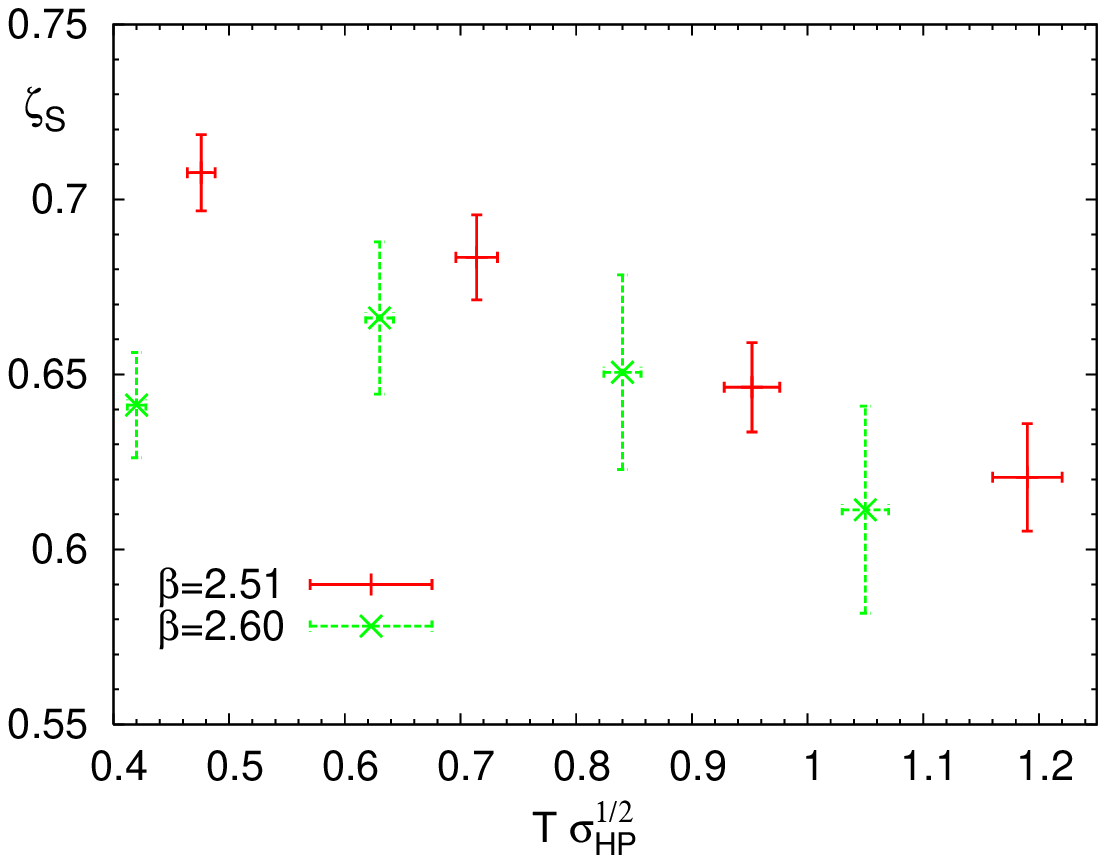} \\
\includegraphics[scale=0.65,clip=false]{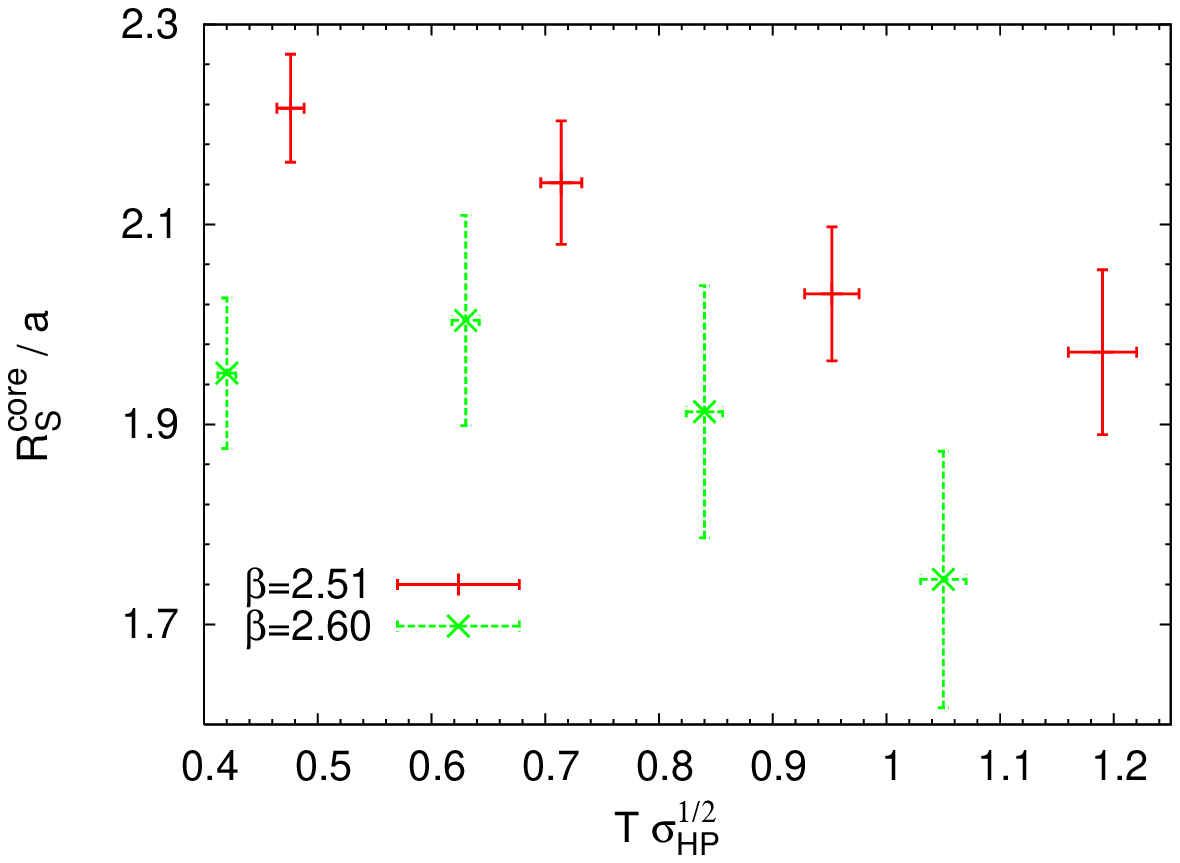} \\[3mm]
\caption{(Color online) The coefficients $\zeta_S$ (top) and $R^{\mathrm{core}}_S/a$
(bottom) of the L\"uscher scaling law~\eq{eq:delta:th} as determined from Wilson loops
of various time extensions $T$  calculated at both  available
data sets. Note that $\zeta_S$
is shown in physical units with the normalization $\sigma = \sigma_\hp$,
while $R^{\mathrm{core}}_S$ is plotted in the lattice units.}
\label{fig:action:alpha-R}
\end{figure}

Next we estimate the parameters $\zeta_S$ and  $R^{\mathrm{core}}_S$
entering the logarithmic scaling law \eq{eq:delta:th}.
To this end, we fitted the $\delta^2_S(R)$ dependence to Eq.~(\ref{eq:delta:th})
in the range $R\sqrt{\sigma_\hp} \geq 0.7$ for $T$ values, for which the finite volume corrections
are expected to be small, namely, for $T\sqrt{\sigma_\hp} < 1.0$  on $\beta=2.51$ configurations and for
$T\sqrt{\sigma_\hp} = 0.42(1), 0.63(1)$ with $\beta=2.60$ data set.
For larger temporal extensions of the Wilson loops we restricted the fitting range from above,
$R\sqrt{\sigma_\hp} \lesssim 1.3$, although we checked that inclusion of largest $R$
points changes the results at most by $\sim 10\%$. The outcome of the fits is presented
on Figure~\ref{fig:action:alpha-R}. As might be expected, the amplitude of the logarithmic
widening seems to be given indeed in physical units,
as is illustrated on the top panel of Figure~\ref{fig:action:alpha-R}.
Namely, if we assume that the string width scale $\sigma$, Eq.~\eq{eq:delta:th},
equals to the projected string tension $\sigma_\hp$ then the corresponding dimensionless
coefficient $\zeta_S$ appears to be spacing independent and is about unity.
However, the data for string core width favors again the constancy
of $R^{core}_S$ in lattice units, the value of which is close to one obtained from the
topological density profiles (see the bottom panel of Figure~\ref{fig:action:alpha-R}).
Note that due to the generically expected small sensitivity of logarithmic fits to $R^{core}_S$,
our statistics and the spacing variations considered might be insufficient to determine accurately
the width of the string core. Hence at present we cannot fairly insist on the relation
$R^{core}_S \sim a$, new measurements are needed to confirm it.

To summarize, the measurements of the action density correlation function \eq{SC}
confirm the logarithmic widening of the confining string. Moreover, we were able to demonstrate
that the string broadening is in agreement with general theoretical expectations and is qualitatively compatible
with the bosonic string picture of the confining flux tube.
However, the data for the string core width, where the string widening settles in,
is less conclusive and favors rather small value of $R^{core}_S$ of order lattice spacing.
In particular, even if one would insist that $R^{core}_S$ is given in physical units,
then its numerical value is about
\beq
\label{R_core}
R^{\mathrm{core}}_S \approx  0.1 ~\mathrm{fm}\,.
\eeq
Note that Eq.~(\ref{R_core}) is in qualitative agreement with the literature.
Indeed, it had been shown in Refs.~\cite{Luscher:2002qv,Juge:2002br}
that the transition from perturbative to stringy description of quark-antiquark pair happens
at rather small distances of order $0.2\div 0.3~\mathrm{fm}$ \cite{Luscher:2002qv}.
The early start of the string regime was also found in
Ref.~\cite{Panero:2005iu} in compact $U(1)$ gauge theory in four space-time dimensions.
It could well be that the early onset of the string fluctuations is a universal feature
of the  theories possessing string excitations.

\section{Discussion and Conclusion}
\label{sec:discussion}

In this paper we investigated the topological and the action density profiles
of the confining string in pure $SU(2)$ Yang-Mills theory on the lattice
using the method of the quaternionic projective $\sigma$-model embedding.
Indeed, the main difficulty in the conventional considerations is
caused by UV dominated zero-point fluctuations, which give identically vanishing but numerically destructive
contribution to the correlation functions of the topological/action density and the confining flux tube.
On the other hand, the $\HP{1}$ $\sigma$-model  embedding and corresponding gauge covariant $\HP{1}$-projection
procedure allows to filter out essentially the only leading power divergences from local observables in question,
leaving intact the subleading and the IR dominated contributions.
At very least this property $\HP{1}$ embedding is known to hold in numerical simulations
and hence allows to obtain
rather accurate results for the topological and action density profiles of the confining string.
It is worth mentioning that the known disadvantages of $\HP{1}$ embedding are expected to play no role
in the problem considered. Indeed, the finite non-locality range of the approach should be inessential
for large enough Wilson loop, which are relevant for the present investigation.
Moreover, our results are in agreement with the existing literature, when the comparison is possible.

Our paper presents the first quantitative investigation of the topological
density distribution in the vicinity of the chromoelectric string. Although it had been known for some
time~\cite{string:topcharge:first,string:topcharge} that confining string suppresses the topological fluctuations,
we were able to investigate the problem in details and to show, in particular, that
the confining flux tube in terms of
the topological density corresponds likely to the fluctuating
string. Moreover, we confirmed the generically expected logarithmic string widening
and  we  attempted also to estimate the characteristic string scales.
Our data seems to indicate that in terms of the topological density the confining flux tube
is rather thin object stretched in between quark-antiquark pair.
Although at present we cannot fairly insist on this picture and more data is obviously needed to
confirm the result, nevertheless, it is rather striking and theoretically challenging  by itself.
We hope to shed more light on the issue in forthcoming publications.

As far as the action density profiles are concerned, the use of the projected action density
allowed us to obtain rather accurate results confirming the string picture. In this case, the
string formation and its logarithmic widening is also established. Moreover, the
characteristic width of the flux tube is shown to be given in physical units and
the features of its broadening seem to be compatible with the effective bosonic string description.

\begin{acknowledgments}
The authors are grateful to the members of ITEP Lattice Group, Institute for Theoretical Physics
of Ka\-na\-za\-wa University, and Research Institute for Information Science and Education of
Hiroshima University for stimulating environment.
The thorough discussions with V.I.~Shevchenko, P.Yu.~Boyko and E.T.~Akhmedov are kindly acknowledged.
The work of F.V.G. was partially supported by the
grants RFBR-05-02-16306a, RFBR-06-02-163069a, RFBR-05-02-17642,
RFBR-0402-16079, RFBR-03-02-16941 and by INTAS YS grant 04-83-3943.
The work of M.N.Ch. was supported by the JSPS grant No. L-06514
and by Grant-in-Aid for Scientific Research by Monbu-kagakusyo, No. 13135216.
\end{acknowledgments}


\end{document}